# On the area under a continuous time Brownian motion till its first-passage time

Michael J. Kearney[1] and Satya N. Majumdar[2]

[1] Advanced Technology Institute, School of Electronics and Physical Sciences, University of Surrey, Guildford, GU2 7XH. United Kingdom

[2] Laboratoire de Physique Théorique et Modèles Statistique, Université Paris-Sud. Bât. 100 91405 Orsay Cedex. France

*Abstract*

The area swept out under a one-dimensional Brownian motion till its first-passage time is analysed using a Fokker-Planck technique. We obtain an exact expression for the area distribution for the zero drift case, and provide various asymptotic results for the non-zero drift case, emphasising the critical nature of the behaviour in the limit of vanishing drift. The results offer important insights into the asymptotic behaviour of a number of discrete models. We also provide a succinct derivation for the distribution of the maximum displacement observed during a first-passage.

PACS: 02.50.-r, 05.10.Gg, 05.40.Jc

## 1. Introduction

Consider a one-dimensional Brownian motion with drift, whose displacement $y(t)$ evolves in continuous time via the Langevin equation,

$$\frac{dy(t)}{dt} = -u_d + \sqrt{D}\xi(t) \tag{1}$$

where $\xi(t)$ is a zero mean white noise source with correlator $\langle \xi(t)\xi(t')\rangle = \delta(t-t')$. The motion starts at $y(t=0) = y_0 > 0$ and we assume $u_d \geq 0$ so that the drift is towards zero, i.e. $y = 0$. Hereafter we set $D = 1$ without loss of generality. We are interested in the probability distribution of two stochastic variables associated with the process: (i) the first-passage time, $t_f$, at which the process crosses zero for the first time and, more importantly, (ii) the *area* swept out by the process till its first-passage time, $A = \int_0^{t_f} y(t')\,dt'$; see figure 1. A motivating factor is to provide insights into the large-scale behaviour of certain lattice polygon models; the results are also of interest in their own right and should find wider application in other fields as well.

We recall that lattice polygons serve as models for a number of interesting physical systems, such as ring polymers, vesicles and percolation clusters [1]. For particular subclasses which are partially directed, including column-convex polygons, bar-graph polygons, staircase polygons etc. [2], it is known that the area-perimeter generating function has a universal scaling form about a tri-critical point (this form being, essentially, the logarithmic derivative of the Airy function) [2, 3]. There is powerful evidence that the same basic scaling form also describes the behaviour of all rooted self-avoiding polygons [4]. It is only for staircase polygons, however, that there exists a rigorous proof of this result starting from the area-perimeter generating function itself, and even then the analysis is far from trivial [5]. We shall show that the scaling behaviour for these partially directed cases is obtained relatively simply from the Brownian motion perspective, wherein the universality aspect follows naturally. The connection is easily made; for example, as described in [6], the two arms of a closed rooted staircase polygon can be thought of as two independent random walks whose trajectories meet for the first time at the end of the polygon. The difference walk between these two arms then defines, in the limit, a Brownian motion. Such a picture also makes a direct connection with the directed Abelian sandpile model proposed in [7], such that $t_f$ relates to the avalanche duration and $A$ to the size of the avalanche cluster [6].

A different area of application arises in queueing theory. Referring to figure 1, one can identify $y(t)$ with the length of a queue at time $t$, whereupon $t_f$ is the busy period (i.e. the time until the queue is first empty) and $A$ is the cumulative waiting time experienced by all the 'customers' during a busy period. A discussion from the discrete perspective may be found in [6]. Another variable of interest in queueing theory is the maximum queue length experienced during a busy period, namely the maximum displacement $y_m = \sup\{y(t) : t \in [0, t_f]\}$ observed till the first-passage time; see figure 1. We return to a discussion of this later.

Existing studies of the integral of the absolute value of Brownian motion over the *fixed* time interval [0,1], i.e. $A = \int_0^1 |y(t')| dt'$ with $y(0) = 0$, have yielded exact results in the limit of zero drift for bridges (which are conditioned on $y(1) = 0$) [8, 9], meanders (conditioned on $y(t) > 0$ for $0 < t < 1$) [10], excursions (conditioned on $y(t) > 0$ for $0 < t < 1$ and $y(1) = 0$) [10-12], as well as more general motions [13, 14]. A recent path integral derivation for meanders and excursions may be found in [15]. The distinguishing features of the present problem are the stochastic nature of the domain of integration, $[0, t_f]$, which prevents a trivial scaling of existing results, and the inclusion of drift, which highlights the critical nature of the process as $u_d \to 0$. We observe that when $u_d > 0$ the process will eventually reach zero with probability one, and is thus *persistent*, whereas when $u_d < 0$ the probability of reaching zero is $1 - e^{-2|u_d|y_0} < 1$ [16]. Since infinite events occur with non-zero probability when $u_d < 0$, the moments of $t_f$, $A$ etc. diverge as $u_d \to 0$ from above. The behaviour for $u_d < 0$, conditioned on the first-passage time being *finite*, is qualitatively the same as the behaviour for $u_d > 0$. Therefore we do not consider the $u_d < 0$ case any further.

## 2. The Fokker-Planck approach

To calculate the distributions of interest, we first consider a general case where we evaluate the probability distribution $P(T, y_0)$ of the observable,

$$T = \int_0^{t_f} V[y(t')] \, dt' \tag{2}$$

where $V[y(t)]$ is an arbitrary functional of the process $y(t)$ and $t_f$ is the first-passage time of the process. If we can obtain the distribution $P(T, y_0)$ for a general $V[y(t)]$, then the different special cases will follow by choosing the functional $V$ properly. For example, if we choose $V[y] = 1$, Eq. (2) gives $T = t_f$ and $P(T, y_0)$ gives the first-passage time distribution. Similarly, the choice $V[y] = y$ provides the distribution of the area $T = A = \int_0^{t_f} y(t') \, dt'$ under the process till it hits zero for the first time. Below we consider general $V[y]$ and derive explicit results for the special cases later.

It is useful to consider the Laplace transform with respect to $T$ of the distribution $P(T, y_0)$,

$$\tilde{P}(s, y_0) \equiv \int_0^\infty P(T, y_0) e^{-sT} \, dT = \left\langle \exp\left\{-s \int_0^{t_f} V[y(t')] \, dt'\right\} \right\rangle \tag{3}$$

where $\langle \ \rangle$ denotes an average over all realisations of the process till the first-passage time. We employ a special backward Fokker-Planck technique that has recently been used in the context of a particle moving in a random Sinai potential [16, 17]. A typical path of the process over the interval $[0, t_f]$ is split into two parts: a left interval $[0, \Delta t]$ where the process proceeds from $y_0$ to $y_0 + \Delta y = y_0 - u_d \Delta t + \xi(0) \Delta t$ in a small time $\Delta t$ and a right interval $[\Delta t, t_f]$ in which the process starts at $y_0 + \Delta y$ at time $\Delta t$ and reaches 0 at time $t_f$. The integral $\int_0^{t_f} V[y(t')] \, dt'$ is similarly split. Since $y(t=0) = y_0$, one gets $\int_0^{\Delta t} V[y(t')] dt' = V[y_0]\Delta t + O(\Delta t^2)$. Then Eq. (3) can be written as,

$$\tilde{P}(s, y_0) = \left\langle \exp\left\{-s\int_0^{t_f} V[y(t')]\, dt'\right\}\right\rangle = \left\langle e^{-sV[y_0]\Delta t} \tilde{P}(s, y_0 + \Delta y)\right\rangle_{\Delta y} \quad (4)$$

where we have used the fact that for the right interval $[\Delta t, t_f]$, the starting position is $y_0 + \Delta y$, which itself is random. The average in the second line of Eq. (4) is over all possible realisations of $\Delta y$. We now substitute $\Delta y = -u_d \Delta t + \xi(0)\Delta t$, expand in powers of $\Delta t$ and average over the noise $\xi(0)$. We use the fact that the noise $\xi(t)$ is delta-correlated, i.e. $\langle \xi(0)^2 \rangle = 1/\Delta t$ as $\Delta t \to 0$. Thus $\langle \Delta y \rangle = -u_d \Delta t + O(\Delta t^2)$ and $\langle \Delta y^2 \rangle = \Delta t + O(\Delta t^2)$, whereupon one obtains by collecting terms of $O(\Delta t)$,

$$\frac{1}{2}\frac{\partial^2 \tilde{P}(s, y_0)}{\partial y_0^2} - u_d \frac{\partial \tilde{P}(s, y_0)}{\partial y_0} - sV[y_0]\tilde{P}(s, y_0) = 0. \quad (5)$$

Note that in deriving Eq. (5) we have *only* required that the noise $\xi(t)$ has zero mean and is delta-correlated, i.e. $\langle \xi(t)\xi(t') \rangle = 0$ for $t \neq t'$, whereas $\langle \xi(t)^2 \rangle = 1/\Delta t$. In other words the detailed nature of the noise distribution is irrelevant. This explains why many different discrete lattice models may be mapped onto the present problem in the appropriate limit.

The differential equation (5) is valid in the range $y_0 \in [0, \infty]$ and satisfies the following boundary conditions. When the initial position $y_0 \to 0$, the first-passage time $t_f$ must also tend to zero. Hence the integral $\int_0^{t_f} V[y(t')]dt'$ vanishes and from the definition in Eq. (3), we get $\tilde{P}(s, y_0 \to 0) \to 1$. When the initial position $y_0 \to \infty$, the first-passage time $t_f \to \infty$, hence the integral $\int_0^{t_f} V[y(t')]dt'$ also diverges in this limit, at least when $V[y]$ is a non-decreasing function of $y$. The definition in Eq. (3)

then gives the boundary condition, $\tilde{P}(s, y_0 \to \infty) \to 0$. For completeness, we also note the normalisation condition, $\tilde{P}(s \to 0, y_0) \to 1$.

**3. Analysis of the special cases**

We now examine the special cases. First, let us consider the first-passage time distribution, for which we need to choose $V[y] = 1$, so that $T = t_f$ from Eq. (2). The solution of the differential equation (5) that satisfies the boundary conditions can be easily obtained,

$$\tilde{P}(s, y_0) = e^{u_d y_0} \exp\left(-y_0 \sqrt{u_d^2 + 2s}\right). \tag{6}$$

Inverting the Laplace transform, we get the required first-passage time distribution, which was also presented in [16],

$$P(t_f, y_0) = \frac{1}{\sqrt{2\pi}} \frac{y_0}{t_f^{3/2}} e^{-(y_0 - u_d t_f)^2 / 2t_f} \tag{7}$$

valid for all $y_0 \geq 0$ and $t_f \geq 0$. When $u_d = 0$ and in the limit $t_f \gg y_0^2$, the distribution has an algebraic tail, $P(t_f) \sim t_f^{-3/2}$, which implies that all moments are infinite. When $u_d > 0$, the moments are finite and may be evaluated exactly as,

$$\left\langle t_f^k \right\rangle = \left(\frac{y_0}{u_d}\right)^k \left(\frac{2}{\pi}\right)^{1/2} (u_d y_0)^{1/2} e^{u_d y_0} K_{k-\frac{1}{2}}(u_d y_0) \tag{8}$$

where $K_\nu(z)$ is a modified Bessel function. In deriving Eq. (8) we have made use of the identity [18],

$$\int_0^\infty x^{-\nu-1} e^{-\alpha x - \beta/x} \, dx = 2\left(\frac{\alpha}{\beta}\right)^{\nu/2} K_\nu\left(2\sqrt{\alpha\beta}\right). \tag{9}$$

Despite its appearance, Eq. (8) reduces to a finite polynomial in $u_d y_0$ for integer $k$; e.g. $\langle t_f \rangle = y_0 / u_d$ and $\langle t_f^2 \rangle = y_0 / u_d^3 + y_0^2 / u_d^2$. Note that when the diffusion is small $\langle t_f^k \rangle \approx (y_0 / u_d)^k$, which is expected since then $y(t) \approx y_0 - u_d t$ so that $t_f \approx y_0 / u_d$ for the dominant paths. One can determine the divergence of the moments as $u_d \to 0$ by using the small $z$ expansion of $K_\nu(z) \sim \Gamma(\nu) 2^{\nu-1} z^{-\nu}$ for $\nu > 0$,

$$\langle t_f^k \rangle \sim y_0 \frac{\Gamma(k - \frac{1}{2}) 2^{k-1}}{\sqrt{\pi}} \frac{1}{u_d^{2k-1}}. \tag{10}$$

The first-passage time distribution in Eq. (7) can, of course, be derived by more conventional methods. The derivation here, however, serves to highlight the ease with which the present method may be applied, and illustrates some of the critical aspects of the problem in the $u_d \to 0$ limit. A simple transformation suffices to capture the perimeter scaling behaviour for staircase polygons, see e.g. [6].

Next we turn to the central question of interest to us, namely the distribution of the area $A = \int_0^{t_f} y(t') dt'$ swept under a Brownian motion till its first-passage time. To evaluate the distribution $P(A, y_0)$, we choose $V[y] = y$ in Eq. (2) so that $T = A$. The second order differential equation (5) with $V[y_0] = y_0$ may be simplified by making the transformation $\tilde{P}(s, y_0) = e^{u_d y_0} \hat{P}(s, y_0)$ to remove the first order derivative. The resulting equation for $\hat{P}(s, y_0)$ is simply the Schrödinger equation for a quantum particle moving in a uniform field. The solution that matches the boundary conditions gives the required result,

$$\tilde{P}(s, y_0) = e^{u_d y_0} \frac{\text{Ai}\left(2^{1/3} s^{1/3} y_0 + \frac{u_d^2}{2^{2/3} s^{2/3}}\right)}{\text{Ai}\left(\frac{u_d^2}{2^{2/3} s^{2/3}}\right)} \tag{11}$$

where $\text{Ai}(z) = \pi^{-1}\sqrt{z/3}\, K_{1/3}(\tfrac{2}{3} z^{3/2})$ is the Airy function [18]. Eq. (11) is our main result. It is evidently more complicated than the corresponding result for the first-passage time distribution given by Eq. (6). To make further progress, we consider the cases $u_d = 0$ and $u_d > 0$ separately.

When $u_d = 0$ one has $\tilde{P}(s, y_0) = 3^{2/3}\Gamma(\tfrac{2}{3})\text{Ai}(2^{1/3} s^{1/3} y_0)$, the inversion of which can be carried out *exactly* using the identity in Eq. (9). This allows us to obtain the distribution $P(A, y_0)$ for all $A \geq 0$ and all $y_0 \geq 0$,

$$P(A, y_0) = \frac{2^{1/3}}{3^{2/3}\Gamma(\tfrac{1}{3})} \frac{y_0}{A^{4/3}} e^{-2y_0^3/9A}; \quad u_d = 0. \tag{12}$$

We are unaware of this result having been written down explicitly and in full before. For $A \gg y_0^3$, the distribution has an algebraic tail,

$$P(A, y_0) \sim \frac{2^{1/3}}{3^{2/3}\Gamma(\tfrac{1}{3})} \frac{y_0}{A^{4/3}}. \tag{13}$$

A derivation of this tail from the staircase polygon perspective is given in [6]. The power law behaviour can be explained using a simple scaling argument (see also [7]). When $u_d = 0$, typically one expects the scaling $y(t) \sim \sqrt{t}$ for large $t$. It follows from the definition $A = \int_0^{t_f} y(t')dt'$ that the area typically scales as $A \sim t_f^{3/2}$ for large $t_f$, and since $P(t_f) \sim t_f^{-3/2}$ from Eq. (7), it follows immediately that $P(A) \sim A^{-4/3}$ for large $A$. Note, however, that this argument does not reproduce the correct amplitude in Eq. (13), not can it reveal the exponentially singular behaviour as $A \to 0$. The result in Eq. (12) is in excellent agreement with the results of a numerical simulation of the process in Eq. (1) with $u_d = 0$, as shown in figure 2.

The exact inversion of Eq. (11) in terms of standard functions appears to be impossible when $u_d > 0$. We therefore focus on deriving: (i) the behaviour of the moments as $u_d \to 0$ and (ii) the tail of the distribution $P(A, y_0)$ for large $A$, which may be deduced from observing how the moments diverge. In general, the moments are given by $\langle A^k \rangle = (-1)^k \partial_s^k \tilde{P}(s, y_0)\big|_{s=0}$. Thus, for example, the mean area is given by $\langle A \rangle = y_0/2u_d^2 + y_0^2/2u_d$. To determine the behaviour of the $k$-th moment as $u_d \to 0$, we simultaneously take the limit $s \to 0$ in Eq. (11) in such a way that the scaling variable $\tau \equiv u_d^2/2^{2/3} s^{2/3}$ remains *fixed* and *large*. Treating $u_d y_0$ as a small expansion parameter one therefore has,

$$\tilde{P}(s, y_0) = 1 + u_d y_0 - \frac{u_d y_0}{\sqrt{\tau}} F(\tau) + O\left((u_d y_0)^2\right) \qquad (14)$$

where $F(\tau) = -\mathrm{Ai}'(\tau)/\mathrm{Ai}(\tau)$ and $F(\tau) > 0$. This is the *exact* scaling form found in the asymptotic analysis of the area-perimeter generating function of staircase polygons. Previous derivations either *assume* a scaling form in conjunction with dominant balance techniques to analyse a functional equation for the generating function [2, 3], or require sophisticated asymptotic methods to analyse a $q$-series representation of the generating function itself [5]. Since $F(\tau) \sim \sum_{k=0}^{k=\infty} a_k \tau^{-(3k-1)/2}$ for large $\tau$, with $a_0 = 1$ [13, 19], Eq. (14) can be viewed as a power series in $s$. The area moments may therefore be determined in straightforward fashion as $u_d \to 0$,

$$\langle A^k \rangle = y_0 (-1)^{k+1} a_k 2^k k! \frac{1}{u_d^{3k-1}} [1 + O(u_d y_0)]. \qquad (15)$$

The coefficients $a_k$ obey a quadratic recurrence relation,

$$a_k = -\frac{1}{2} \sum_{i=1}^{k-1} a_i a_{k-i} - \left(\frac{3k}{4} - 1\right) a_{k-1} \qquad (16)$$

for all $k \geq 2$ with $a_1 = 1/4$. We note that this recurrence relation also occurs in studies of extremal distributions [11, 15, 20] and in various other enumeration problems in computational science and graph theory [19]. Although an explicit expression for $a_k$ is unknown, its asymptotic behaviour as $k \to \infty$ is known so that as $u_d \to 0$ and $k \to \infty$,

$$\langle A^k \rangle \sim y_0 \sqrt{\frac{2}{\pi}} \left(\frac{3}{8}\right)^k \Gamma\left(2k + \frac{1}{2}\right) \frac{1}{u_d^{3k-1}}. \tag{17}$$

This result is sufficient to determine the corresponding tail of the area distribution $P(A, y_0)$ as $A \to \infty$, which is given by,

$$P(A, y_0) \sim \frac{y_0}{\sqrt{\pi}} \left(\frac{2}{3}\right)^{1/4} u_d^{7/4} A^{-3/4} \exp\left\{-\left(\frac{8}{3}\right)^{1/2} u_d^{3/2} A^{1/2}\right\}. \tag{18}$$

It is readily verified that Eq. (18) implies Eq. (17). That Eq. (18) is, in an appropriate sense, 'unique' follows from the fact that any candidate distribution whose moment sequence satisfies Eq. (17) is uniquely determined by its moment sequence, which follows from the well-known Carleman criterion, with the corollary that all such distributions must be asymptotically equivalent as $A \to \infty$ [6]. The prefactor $A^{-3/4}$ in Eq. (18) was also recently identified in [21] for staircase polygons and column-convex polygons of fixed area with perimeter dependent weights. Technically, Eq. (18) is only valid for $u_d y_0$ small, but it is a salient result nonetheless. The direct inversion of Eq. (11) might offer the prospect of extending beyond this regime, but the technical difficulties are formidable.

**4. Characterising the maximum displacement**

We now return to the question posed earlier relating to the distribution of the maximum displacement $y_m$ observed till the first-passage time; see figure 1. The particular Fokker-Planck technique we employ may be used to derive this succinctly and elegantly. The equivalent problem for the Edwards-Wilkinson fluctuating interface model has been solved recently and provides a rare example of an exact distribution for the maximum of a set of strongly correlated random variables [15, 20]. The present problem may be viewed similarly. To proceed, imagine summing over all paths for which the maximum displacement $Y \leq y_m$, by first stepping from $y_0$ to $y_0 + \Delta y$, followed by again summing over all paths for which $Y \leq y_m$. In this way, one can relate the probability $\Pr(Y < y_m | y_0)$ to itself under all possible evolutions with respect to the initial condition $y_0$. The Markov nature of the process means that $\Pr(Y < y_m | y_0) = \langle \Pr(Y < y_m | y_0 + \Delta y) \rangle_{\Delta y}$, where the average $\langle \; \rangle_{\Delta y}$ defined in Eq. (4) effects the sum over all the weighted possibilities of getting from $y_0$ to $y_0 + \Delta y$. Expanding as before and averaging gives,

$$\left[ \frac{1}{2} \frac{\partial^2}{\partial y_0^2} - u_d \frac{\partial}{\partial y_0} \right] \Pr(Y < y_m | y_0) = 0 \tag{19}$$

with boundary conditions $\Pr(Y < y_m | y_0 = 0) = 1$ and $\Pr(Y < y_m | y_0 = y_m) = 0$. The solution is,

$$\Pr(Y < y_m | y_0) = e^{u_d y_0} \frac{\sinh(u_d (y_m - y_0))}{\sinh(u_d y_m)}; \quad y_0 \leq y_m. \tag{20}$$

This is the probability that a one-dimensional random walker on the interval $[0, y_m]$ first exits via the boundary at $y = 0$. The required distribution may be obtained in a straightforward manner by evaluating $P(y_m, y_0) \equiv (d/dy_m) \Pr(Y < y_m | y_0)$,

$$P(y_m, y_0) = u_d e^{u_d y_0} \frac{\sinh(u_d y_0)}{\sinh^2(u_d y_m)}; \quad y_m \geq y_0. \tag{21}$$

When $u_d = 0$, $P(y_m, y_0) = y_0 / y_m^2$ for $y_m \geq y_0$ and all the moments are infinite. The maximum displacement for zero drift therefore has a power law distribution (inverse square law), which can be justified as before using a simple scaling argument. Thus, typically, $y_m \sim t_f^{1/2}$, and since $P(t_f) \sim t_f^{-3/2}$, it follows that $P(y_m) \sim y_m^{-2}$. When $u_d > 0$ the moments are finite and, in particular, the mean value is given by,

$$\langle y_m \rangle = y_0 + \frac{e^{2u_d y_0} - 1}{2u_d} \log\left(\frac{1}{1 - e^{-2u_d y_0}}\right). \tag{22}$$

The logarithmic divergence as $u_d \to 0$ is somewhat unexpected and should be contrasted with the power law divergence of the mean first-passage time $\langle t_f \rangle \sim u_d^{-1}$ and the mean area $\langle A \rangle \sim u_d^{-2}$. We note that $\langle y_m \rangle \langle t_f \rangle < \langle A \rangle < \langle y_m t_f \rangle$ as $u_d \to 0$, where the last inequality follows because $A < y_m t_f$ with probability one. Thus the covariance, $\text{cov}(y_m, t_f)$, diverges strongly as $u_d \to 0$, implying a close correlation between $y_m$ and $t_f$. An exact derivation would be of interest. It is also worth noting that Eq. (21) and Eq. (22) establish the behaviour of the maximum avalanche width in the canonical directed Abelian sandpile model [6, 7]. We are unaware of such results having been presented in this context before.

## 5. Summary

Using a special backward Fokker-Planck technique, the distributions of various stochastic variables associated with a one-dimensional Brownian motion till its first-passage time have been analysed. The results help to explain the asymptotic behaviour of various discrete lattice models which arise naturally in statistical physics and queueing theory. Regarding the first-passage time and maximum displacement distributions, the results are complete. Regarding the area distribution, we have solved the zero drift case exactly, and presented asymptotic results for when the drift is non-zero. The remaining outstanding challenge is to invert Eq. (11) exactly.

**References**

[1] van Rensburg E J J 2000 *The statistical mechanics of interacting walks, polygons, animals and vesicles* (Oxford: Oxford University Press)

[2] Prellberg T and Brak R 1995 *J. Stat. Phys.* **78** 701

[3] Richard C 2002 *J. Stat. Phys.* **108** 459

[4] Richard C, Guttmann A J and Jensen I 2001 *J. Phys. A: Math. Gen.* **34** L495

[5] Prellberg T 1995 *J. Phys. A: Math. Gen.* **28** 1289

[6] Kearney M J 2004 *J. Phys. A: Math. Gen.* **37** 8421

[7] Dhar D and Ramaswamy R 1989 *Phys. Rev. Lett.* **63** 1659

[8] Shepp L A 1982 *Annals of Probab.* **10** 234

[9] Takacs L 1992 *Annals of Appl. Probab.* **2** 435

[10] Takacs L 1995 *J. Appl. Probab.* **32** 375

[11] Darling D A 1983 *Annals of Probab.* **11** 803

[12] Louchard G 1984 *J. Appl. Probab.* **21** 479

[13] Takacs L 1993 *Annals of Appl. Probab.* **3** 186

[14] Perman M and Wellner J A 1996 *Annals Appl. Probab.* **6** 1091

[15] Majumdar S N and Comtet A to be published; arXiv: cond-mat/0409566

[16] Majumdar S N and Comtet A 2002 *Phys. Rev. E.* **66** 061105

[17] Dean D S and Majumdar S N 2001 *J. Phys. A: Math. Gen.* **34** L697

[18] Gradshteyn I S and Ryzhik I M 1980 *Tables of Integrals, Series and Products,* 5$^{th}$ ed. (London: Academic Press)

[19] Flajolet P and Louchard G 2001 *Algorithmica* **31** 361

[20] Majumdar S N and Comtet A 2004 *Phys. Rev. Lett.* **92** 225501

[21] Rajesh R and Dhar D *Phys. Rev. E.* (to appear); arXiv: cond-mat/0303577

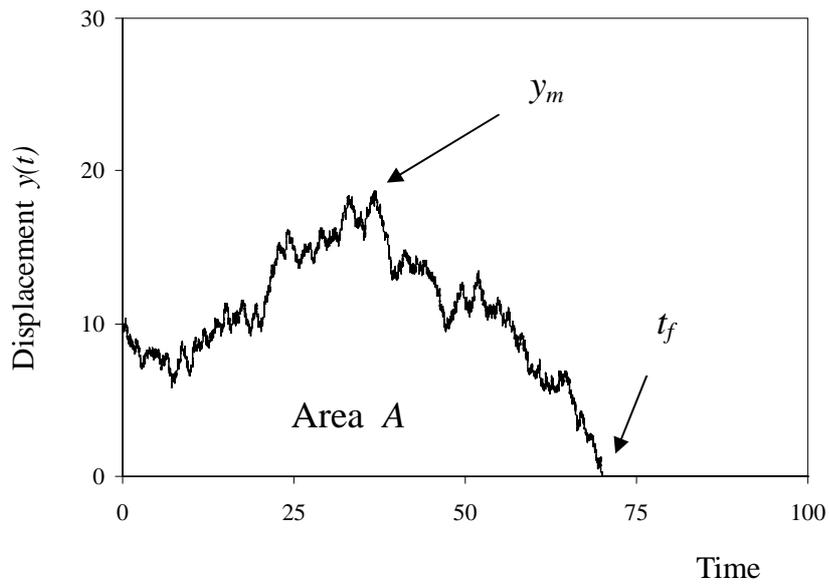

**Figure 1.** Sample of the process $y(t)$ for $D=1$ and $u_d = 0.05$, with initial condition $y_0 = 10$. The first passage time $t_f \approx 70$, whilst the area under the process up to the first passage time $A \approx 760$. The maximum $y_m \approx 18.7$.

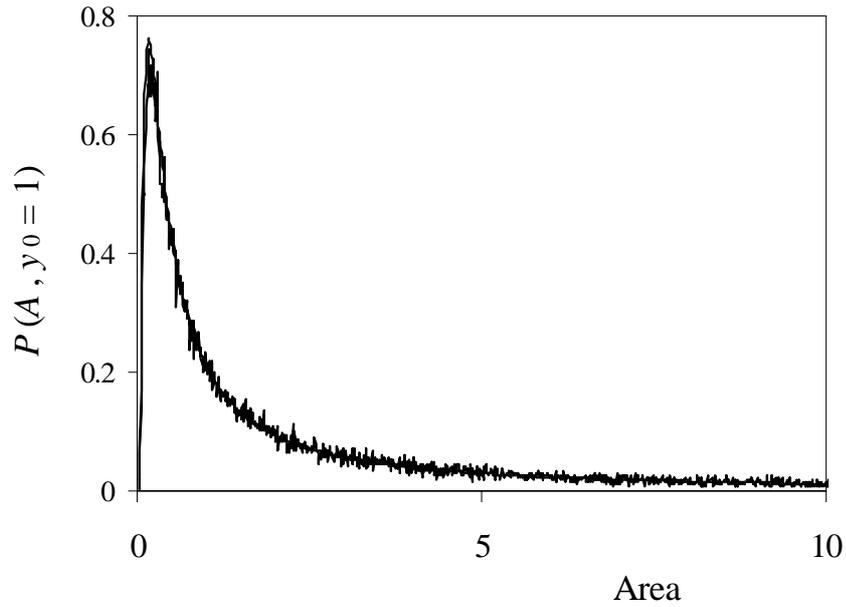

**Figure 2.** The zero drift distribution $P(A, y_0)$ obtained numerically by simulating the process $y(t)$ in Eq. (1) with $y_0 = 1$ compared with the analytical expression in Eq. (12) (superimposed solid line). The numerical histogram is obtained from $10^5$ samples with a bin size of 0.01. As presented, both curves are normalised with respect to a cut-off area $A^* = 100$, following the procedure described in [17].